# Simple Analytical Model for Optimizing Integrating Sphere Port Sizes

A.M. Bratkovsky [1,2,*]

[1] Kapitza Institute for Physics Problems, 2 Kosygina Str., 119334 Moscow, Russia

[2] Corning R&D Corp., 1 Riverfront Plaza, Corning, NY 14831, USA

**Abstract**

The integrating sphere (IS) is an indispensable tool for measuring transmission and scattering of materials and their colorimetry, as well as other photometric tasks. The accuracy of its data depends critically on port sizes used for measurement and control, usually defined by trial and error or brute-force optical simulations. To find the optimal port sizes of this powerful tool, a *sample visibility* function is defined and optimized using the energy conservation principle. This yields an analytical expression that should be useful in a variety of applications, especially those where signal is rather small (low-haze materials).

**Introduction**

The Integrating Sphere (IS) is a powerful tool used extensively in photometry and colorimetry invented by R. Ulbricht over a century ago [1,2]. We shall recall below the main features of this ingenious device that is now available from multiple commercial sources for use in photometry, colorimetry, haze, and other measurements [2]. The applications area grows to this day, now including illumination [3] and communication [4]. In spite of the apparent simplicity of the device, theoretical understanding of its operation was slowly developing over decades with multiple approaches applied (cf. Ref. [6]). Still, issues related to optimal configuration like the port sizes are left to the 'rule of thumb' [5,7] or direct simulation (see [8-10] and references therein). One important use of IS is characterization of diffuse scattering from transparent low-haze samples. In this, as well as all other cases, it would help to have a simple analytical formula for an optimal sample port size. Such a formula is derived below from the energy conservation principle.

**Irradiance and sample 'visibility' in integrating sphere**

Consider a radiation exchange between small surface elements dA and dA', Fig.1(a). If the element dA' has radiance $L$ [W.m$^{-2}$.sr$^{-1}$], i.e. it radiates $L$ Watts per unit area into a unit solid angle (steradian, sr) then the element dA a distance $s$ away from dA' will receive energy of *irradiance* E(P) [W.m$^{-2}$] times dA:

$$E(P)dA = L(P' \to P)dA' \cdot d\Omega_s = L(P' \to P)dA' \cdot \frac{dA \cos\vartheta}{s^2} \quad [W] \qquad (1)$$

One has $\vartheta' = \vartheta$, $s = 2R\cos\vartheta$ for a sphere [see Fig. 1(b). Assuming the element dA' on such a sphere is the Lambertian source with $L(P' \to P) = L\cos\vartheta'$, it will receive the power

$$E(P)dA = L\cos\vartheta' \, dA' \cdot \frac{dA\cos\vartheta}{s^2} = \frac{L \, dA' dA}{4R^2} = \frac{\pi L}{A_s} dA' dA \quad [W] \qquad (2)$$

Here, $A_s = 4\pi R^2$ is the surface area of the sphere. One sees that sphere with diffuse coating has an amazing property that its every radiant element produces the same irradiance at all elements of the sphere regardless of their relative positions. Therefore, the first ray hitting the sphere would spread its luminous energy evenly over the entire sphere and every subsequent ray would too. A sphere is likely to be the only surface with a property $s^{-2} \cos\vartheta' \cos\vartheta =$const, Fig.1(b).

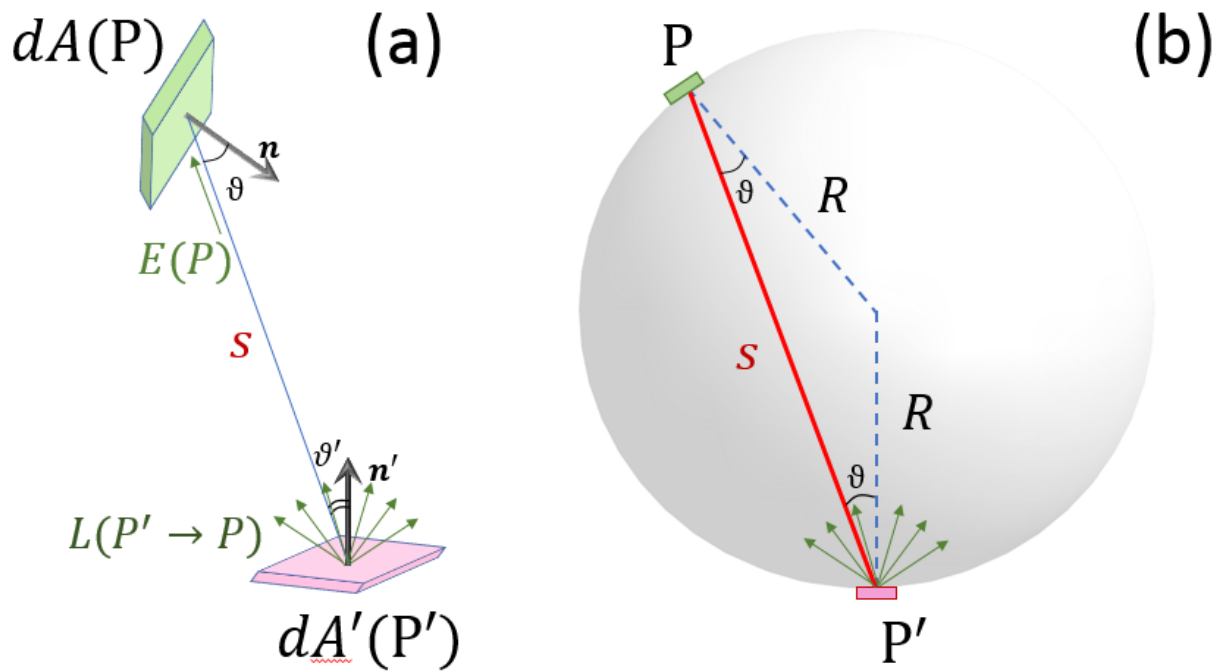

**Fig.1.** (a) Schematic of the radiation power exchange between two Lambertian (ideally light diffusing) elements, (b) only in a sphere, every radiant element produces the *same* irradiance at all elements of the sphere regardless of their relative positions (see text.) Cumulative irradiance of every element of the sphere after multiple scattering events would much exceed the very first irradiance of the wall by inserted light before it receives additional light scattered by the sphere.

The sphere could be used in both reflection (Fig.2) and transmission configurations. The fast low-haze sample characterization requires optimizing the collection of light diffusely scattered by the sample. The low-haze samples would find very wide usage in e.g. mobile devices. The question is then: what is the port size for a sample window that maximizes the reading by the detector?

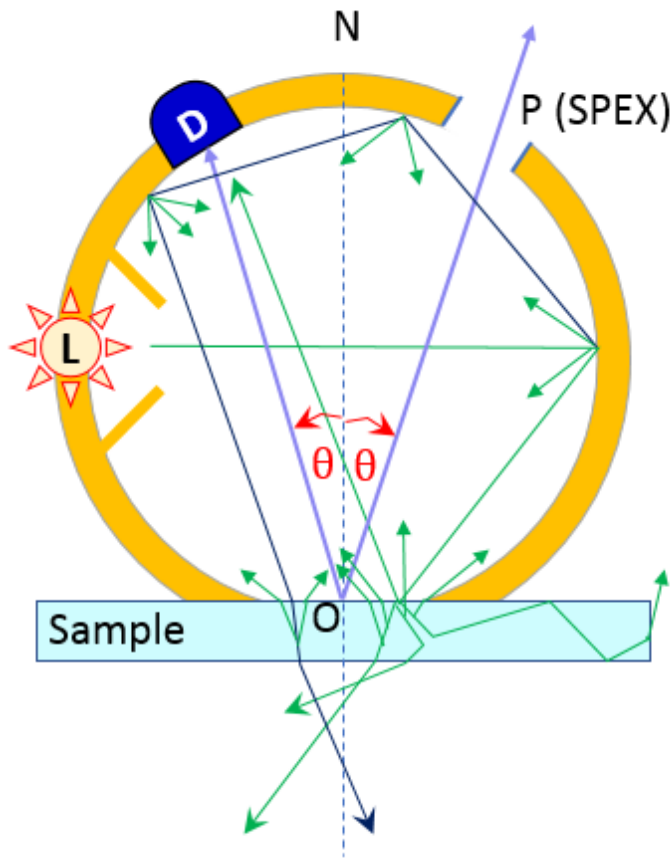

**Fig.2**. Schematic of the Integrating sphere setup for measuring diffuse scattering from sample in specular light *excluded* (SPEX) geometry. The baffled light source L introduces power $P$ into the sphere that diffusely scatters off the walls with high *reflectance* $m_w$ close to unity, with the rest $(1 - m_w)$ absorbed. The sample gets homogeneously illuminated from top half space and the light diffusely scattered by the sample gets collected by the detector D. The detector port diameter is $d_D$. Open port P with the same diameter removes light specularly reflected from the sample. The detector and SPEX ports are positioned symmetrically with respect to the North pole of the sphere.

To answer this question, we apply the standard theory of Integrating Sphere (IS) [2]. We shall use the following notations with comments:

- $P$, the total energy inserted by light source into the IS
- Sphere has $N$ ports occupying areas $A_i = f_i A$, where $A = 4\pi R^2$ is the total area of the sphere with radius $R$, $D = 2R$ its diameter, $i = 0 \div N$. Here, $f_i$ is the *fractional area* occupied by the $i$th port.
- 'Port' $i = 0$ is the area first hit by the light from the source. In the present case, its reflectance is therefore the same as the wall, $m_0 = m_w$ (Fig.2.) The spherical geometry guarantees that the initial beam will scatter diffusely like all subsequent and *every* scattering event will distribute power uniformly over the whole sphere. This fact is of *cardinal importance* for the use of the Integrating Spheres in photometry.

- Ports have total (integrated over all angles back into the IS) reflectances $m_i$. For transparent samples, $m_s \ll 1$ is the 'scatter ratio' for reflection. We will use $m_s = 0.0025$ as an example.
- Steady power hitting the $i$th port is $P_i = f_i P$.
- Sphere walls have reflectance $m_w$. We shall use a typical $m_w = 0.98$ (Spectralon). The wall absorbance is obviously small, $1 - m_w \ll 1$.
- Fraction of the surface occupied by the detector and the SPEX port is $2f_d$. (explained in the caption to Fig.2). We shall use, for illustrative purposes, $f_d = 0.002504$ (corresponding ratio of port diameter over diameter of the sphere $d_D/D = 0.1$) and $f_d = 0.005657$ corresponding to $d_D/D = 0.15$. We assume that the detector reflectance is small, $m_d \ll 1$, i.e. it absorbs most of the light hitting it.

Apply energy conservation to find *steady flux F* [W/m²] on the surface of the sphere after multiple reflections. Power $P$ [W] inserted into the sphere illuminates area $A_0$ [m²], Fig.2, and a fraction $m_0$ of that power gets reflected diffusely and uniformly illuminates the entire sphere interior with flux $F_0 = m_0 P/A$ [W/m²]. Note that the ports remove power

$$F \sum_{i=0}^{N} (1 - m_i) A_i = FA \sum_{i=0}^{N} (1 - m_i) f_i. \tag{3}$$

The walls absorb power

$$F(1 - m_w)\left(A - \sum_{i=0}^{N} A_i\right) = FA(1 - m_w)\left(1 - \sum_{i=0}^{N} f_i\right). \tag{4}$$

Balance of power yields

$$m_0 P = FA \sum_{i=0}^{N} (1 - m_i) f_i + FA(1 - m_w)\left(1 - \sum_{i=0}^{N} f_i\right), \tag{5}$$

Hence, the average steady flux is equal to

$$F = M \frac{P}{A}, \tag{6}$$

where

$$M = \frac{m_0}{1 - \bar{m}} \gg 1, \tag{7}$$

is the sphere multiplication factor and we have defined the average reflectance,

$$\bar{m} = m_w \left(1 - \sum_{i=0}^{N} f_i\right) + \sum_{i=0}^{N} m_i f_i. \tag{8}$$

Notice that $F \gg F_0$, the latter being the light power flux inserted by the lamp (Fig.2.)

The fraction of steady power (after multiple reflections) hitting port $i$ is, from a total energy balance

$$p_i = \frac{P_i}{P} = \frac{FA_i}{P} = M f_i \gg f_i. \tag{9}$$

Now, we could find a fraction of power hitting the detector D in the situation shown in Fig. 2 (i.e. having three openings and remembering that in our case $m_0 = m_w$ and $m_d \approx 0$) as

$$p_d = \frac{P_d}{P} = M f_d = \frac{m_w f_d}{1 - m_w(1 - 2f_d) + (m_w - m_s) f_s}, \tag{10}$$

where

$$\bar{m} = m_w(1 - f_o - 2f_d - f_s) + m_o f_o + 2m_a f_a + m_s f_s$$
$$\approx 1 - m_w(1 - 2f_d) + (m_w - m_s)f_s \qquad (11)$$

Now, introduce *Sample Visibility V* of the sample scattering *by the Detector*, which is a difference in Detector (D) reading with sample off (the sample port is a hole reflecting very little back into the IS) and with sample on:

$$V = p_d(m_s) - p_d(m_s = 0)$$

$$= \frac{m_w f_d m_s f_s}{[1 - m_w(1 - 2f_d - f_s)][1 - m_w(1 - 2f_d - f_s) - f_s m_s]}$$

$$\approx \frac{m_w f_d m_s f_s}{[1 - m_w(1 - 2f_d - f_s)]^2}, \qquad (12)$$

since $f_s m_s \ll 1$. The *sample visibility* increases linearly with $f_s$ with a large prefactor as $\frac{m_w f_d m_s f_s}{(1 - m_w)^2} \sim 10^4 \, f_d m_s f_s$ and has a peak as a function of the sample fractional area at

$$f_{s,max} = \frac{1 - m_w}{m_w} + 2f_d, \qquad (13)$$

see Fig.3. Above, the term $2f_d$ is the fractional area occupied by the open ports. More generally, the *optimal* fractional area of a sample port is

$$f_{s,max} = \frac{1 - m_w}{m_w} + \Sigma f_{\text{open ports}}. \qquad (14)$$

The above Eq.(14) is the main result of the paper.

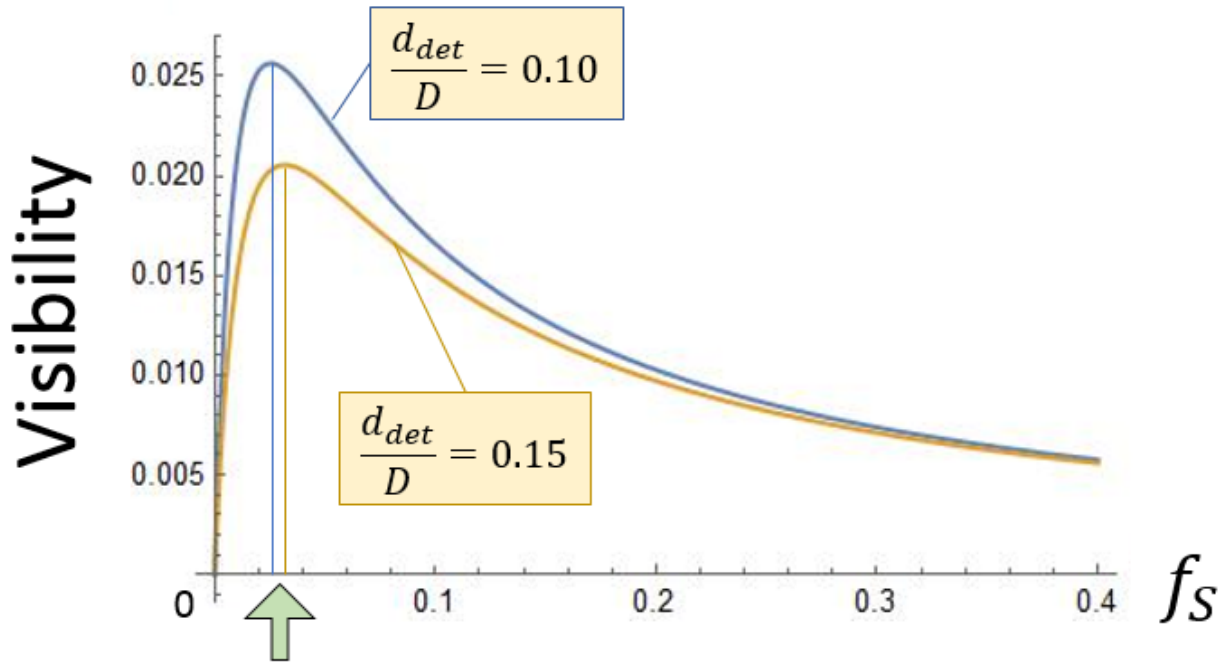

Fig.3. The *Sample Visibility* V of the diffuse scattering of the sample versus the fractional area of the sphere occupied by the sample port $f_s$. Arrow marks the optimal visibility that is reached for the fraction of sample area versus sphere area $f_{S,max} = 0.025 - 0.030$ (i.e. $2.5 - 3\%$), when the ratio of sample port diameter $d_{\text{sample}}$ to the sphere diameter $D$ is $\frac{d_{\text{Sample}}}{D} = \frac{5}{16} \div \frac{1}{3}$.

Apparently, the sample visibility peak position at $\frac{d_{\text{Sample}}}{D} \approx \frac{1}{3}$ is not very sensitive to the detector fractional size. In fact, it slightly increases for a smaller detector size. This will reduce the absolute amount of power registered by the Detector, so further optimization step might account for threshold power required by the given detector at a given power of the light source L, Fig.2.

**Conclusions**

The Integrating Sphere is a simple yet powerful tool for measuring scattering power by transparent materials, diffusers, etc. It could be used for fast characterization of important class of low-haze samples and its sensitivity could be significantly optimized by using sample port size that occupies about $f_S = 2.5 - 3\%$ of the sphere surface area, i.e. the sample port diameter being about one third of the Sphere diameter. Further optimization of the tool is possible with account of threshold power on the detector. The above analysis should help with designing new spheres without customarily used 'rule of thumb' [5-7] or numerical optimization via ray tracing simulations. It is interesting to note that the commercially available spheres feature port sizes that are pretty close to those predicted above (Fig.3) [11], complying with empirical 'rule of thumb' that the ports occupy no more than 5% of sphere surface [5,7].

The author would like to acknowledge useful comments by Dr. Michal Mlejnek.

**References**

* alex.bratkovski@gmail.com

[1] https://en.wikipedia.org/wiki/Integrating_sphere (accessed on Sep 25, 2023). The practical implementation of the integrating sphere was due to work by R. Ulbricht published in 1900.

[2] E. Karrer, The use of the Ulbricht sphere in measuring reflection and transmission factors, J. Opt. Soc. Amer. **5**, 96 (1921); https://doi.org/10.1364/JOSA.5.000096.

[3] A. Ducharme, *et al*. (1997). Design of an Integrating Sphere as a Uniform Illumination Source. IEEE Trans. Ed. **40**, 131 (1997).

[4] A. Dey, *et al*., Design of High-Performance LED Based Integrating Sphere for Illumination and Communication, Intl. Conf. Opto-Elec. Appl. Optics (Optronix), Dey2019OH, pp. 1-6 (2019).

[5] D.G. Goebel, Generalized integrating-sphere theory, Appl. Optics **6**, 125 (1967).

[6] J.F. Clare, Comparison of four analytic methods for the calculation of irradiance in integrating sphere, J. Opt. Soc. Am. A **15**, 3086 (1998).

[7] K.F. Carr, Integrating sphere theory and applications, Surf. Coatings Intl. **8**, 380 (1997).

[8] B.G. Crowther, Computer modeling of integrating spheres, Appl. Opt. **35**, 5880 (1996).

[9] A.V. Prokhorov, S. N. Mekhontsev, and L. M. Hanssen, Monte Carlo modeling of an integrating sphere reflectometer, Appl. Opt. **35,** 3832 (2003).

[10] C. Tang, M. Meyer, B. L. Darby, B. Auguie, and E. C. Le Ru, Realistic ports in integrating spheres: reflectance, transmittance, and angular redirection, Appl. Opt. **57**, 1581 (2018).

[11] See e.g. Ci Systems http://www.ci-systems.com/integrating-sphere , Labsphere http://www.labsphere.com, Newport http://www.newport.com , Konica Minolta https://sensing.konicaminolta.us/us/technologies/integrating-spheres/ (all accessed Sep 25, 2023.)